\documentclass[journal]{IEEEtran}
\usepackage{amsmath, amssymb}
\usepackage{graphicx}
\usepackage{cite}
\usepackage{booktabs}
\usepackage{multirow}
\usepackage{enumitem}
\usepackage{threeparttable}
\usepackage{algorithm}
\usepackage{algorithmic}
\usepackage{xcolor}
\usepackage[a4paper,margin=0.75in]{geometry}
\usepackage{cleveref}
\usepackage{url}
\usepackage{subcaption}

\newcommand{\Iset}{\mathcal{I}}

\begin{document}

\title{TRUST: Item-Calibrated Interval Evidence for Temporal Session-Based Recommendation}

\author{
    Linjiang Guo, Nitin Bisht, 
    Shiqing Wu, Yifan Yin, and Guandong Xu\textsuperscript{*}
    \thanks{
        Linjiang Guo, Nitin Bisht, and Yifan Yin are with the School of Computer Science, University of Technology Sydney, Australia. E-mails: linjiang.guo@student.uts.edu.au, nitin.bisht@student.uts.edu.au,
        Yifan.Yin-2@student.uts.edu.au.
    }
    \thanks{
        Shiqing Wu is with the Faculty of Data Science, City University of Macau, Macau SAR. E-mail: sqwu@cityu.edu.mo.
    }
    \thanks{
        Guandong Xu is with The Education University of Hong Kong, Hong Kong SAR. E-mail: gdxu@eduhk.hk.
    }
    \thanks{\textasteriskcentered Corresponding authors.}
}

\maketitle

\begin{abstract}

Temporal signals have been widely used in session-based recommendation to infer user interest. Existing temporal session-based recommenders primarily rely on absolute interval values, implicitly assuming that the same interval carries similar interest signals across items. However, we empirically find that this assumption does not hold: each item has its own interval distribution, so an interval should be interpreted relative to the item it belongs to. Based on this observation, we propose TRUST, a framework that evaluates each observed interval relative to the empirical interval distribution of the corresponding item. Specifically, we propose a score function to guide global neighbor sampling, session graph encoding, and final interest aggregation. Experiments on public datasets show that TRUST consistently improves over representative temporal and non-temporal baselines, and plug-in experiments further show that the proposed scoring function can improve existing temporal session recommenders as a model-agnostic method. Component-wise ablations further show that calibrating the temporal signals within each module, rather than removing the module itself, consistently improves neighbor sampling, session graph encoding, and interest aggregation.
\end{abstract}

\section{Introduction}

Temporal session-based recommendations (TSBRs) aim to predict a user's next interaction by modeling both the order and timing of interactions within an ongoing session. Unlike conventional session-based recommendation (SBR), which mainly relies on item co-occurrence and transition order\cite{li2024graph, choi2025linear,zhang2026rethinking}, TSBRs further use temporal signals as user-interest evidence for estimating the user's current interest \cite{yi2014beyond,bogina2017incorporating,xu2025time,guo2026dual}. For example, if a user’s interaction with an item is consistent with common interval patterns, this may indicate stronger interest. In contrast, interactions that fall substantially below typical interval patterns, such as very short interval time, may suggest weaker or even negative interest \cite{yin2013silence, gong2022positive,xie2023reweighting,he2023survey}. 

\begin{figure}[t]

  \centering
    \includegraphics[width=\linewidth]{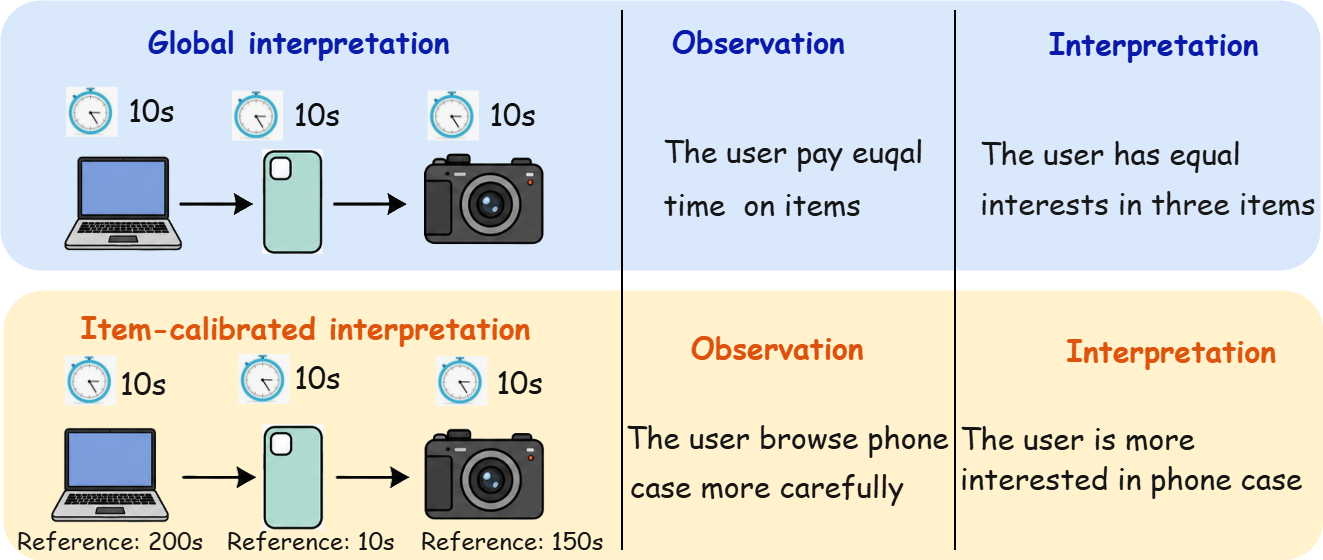}
  \caption{%
Motivation for item-calibrated temporal interpretation. 
The same observed interval can carry different user-interest evidence when evaluated against item-specific temporal references.
  }\label{fig:toy_example}
  \label{fig:motivation}
\end{figure}

Existing TSBR methods usually interpret such intervals according to a global interval distribution over all training interactions. Prior analyses show that intervals often cluster around a typical range and become sparse at very short or very long durations (as Overall label shown in \Cref{fig:engagement-violin}) \cite{wu2022feedrec,guo2026dual}. This observation motivates TSBRs to weight session interactions by their temporal behavior: interactions with typical intervals are treated as stronger evidence of user interest, while unusually short or long ones indicate less interest \cite{kim2014modeling}. However, users do not naturally spend the same amount of time on every item, even when they are interested \cite{liu2010understanding,seki2018analysis,xie2023reweighting}. As illustrated in \Cref{fig:toy_example}, a user spends 10 seconds on a laptop, a phone case, and a camera. Existing TSBRs treat them as the same evidence of user interest because the intervals are the same. When item-specific typical intervals (reference interval) are considered, the interpretation changes: the reference interval is 200 seconds for the laptop, 10 seconds for the phone case, and 150 seconds for the camera. Relative to these references, the phone-case interaction is temporally typical, whereas the laptop and camera interactions are abnormal. Thus, intervals should be interpreted relative to the item being viewed, rather than only by their absolute value. We refer to this item-relative consistency between an observed interval and the item's usual interval distribution as temporal reliability.


This motivates our main objective: to make temporal intervals reliable evidence for temporal session-based recommendation by interpreting each interval relative to the item being viewed, rather than treating the same interval as equally meaningful across all items. Realizing this objective is not straightforward: (1) how to verify whether item-level interval heterogeneity makes globally interpreted intervals unreliable for TSBR; (2) how to estimate item-calibrated temporal signal without becoming overconfident for sparsely observed items; and (3) how to use this reliability signal in global relation modeling, local session transition modeling, and final interest generation.
We therefore propose a framework named \textit{\textbf{T}emporal \textbf{R}eliability \textbf{U}nder item-\textbf{S}pecific in\textbf{T}erval evidence for Session-based Recommendation} (TRUST), which addresses these challenges with three corresponding designs. First, we provide empirical evidence through distributional and perturbation analyses in \Cref{sec:stat}, showing that item-specific interval patterns differ from the overall interval distribution and that this mismatch affects performance of TSBRs. Second, we propose a score function (ITSF) to calibrate the heterogeneity of item-specific interval distribution: ITSF estimates an interval's reliability score from the interval's empirical quantile within the viewed item-specific interval distribution, and relaxes the penalty when that distribution is estimated from few observations. Third, we use the ITSF to guide the neighbor sampling, graph learning, and generation of session representation in TRUST. Specifically, in the global graph, TRUST uses transitions with reliable intervals to build item neighborhoods, reducing the influence of noisy co-occurrences on collaborative signals. Within each session, TRUST helps the graph encoder distinguish transitions based on how well their intervals align with the corresponding items' interval distributions. Finally, TRUST uses interval reliability to decide whether to focus on recent interactions or look back to earlier ones. The overall framework is depicted in \Cref{fig:framework}. We summarize main contributions as follows:

\begin{itemize}
\item To the best of our knowledge, we are the first to identify and empirically characterize the impact of uncalibrated intervals on the performance of existing TSBR methods. We show that interval observations are not comparable across items, and that uncalibrated intervals can lead to a drop in performance.

\item We propose an item-specific interval calibration score function that measures how well an observed interval matches the historical interval distribution of the corresponding item. The score function is model-agnostic and can be integrated into existing TSBRs.

\item We develop TRUST, a temporal reliability-guided session recommendation framework that uses the item-calibrated temporal signal to improve both collaborative evidence modeling and session interest learning.

\item Experiments on public datasets show that TRUST improves over strong temporal and non-temporal baselines. Plug-in experiments further demonstrate the model-agnostic applicability of the proposed score function, and ablation studies verify the contribution of each temporal reliability-guided mechanism within the component.
\end{itemize}

\section{Related Work}
\label{sec:related-work}

Early recommender systems mainly infer user interests from historical user--item interactions and collaborative patterns, where item co-occurrence and latent user--item relations are used as the primary evidence for recommendation \cite{koren2009matrix,rendle2009bpr}. In session-based recommendation, the task becomes more challenging because the model usually observes only a short anonymous interaction sequence. Existing session-based methods have therefore developed recurrent \cite{hidasi2015session}, attention-based \cite{li2017neural,wang2026category}, and graph-based architectures \cite{wu2019session,yu2020tagnn} to model interests. These methods have substantially improved the representation of item dependencies within sessions, but most of them treat an interaction mainly as an item identity or an ordered transition. As a result, the temporal evidence associated with each interaction is ignored. This limitation has motivated increasing attention to temporal recommendation models, which incorporate time-related signals to better characterize user interest.

Building on this, temporal recommendations further enhance recommendation by incorporating temporal signals to better capture user's interest \cite{koren2009collaborative, ye2020time,li2024graph,zhang2025survey}. In these scenarios, timestamps \cite{fan2021continuous,tran2023attention,lee2025enhancing}, interaction intervals \cite{ma2020temporal,li2020time,luo2026time}, and temporal ordering \cite{wan2023spatio}, have been used to model preference drift \cite{li2022time,heryawan2025trust}, recency effects \cite{filipovic2021modeling}, and sequential dependency \cite{zhang2021knowledge,xia2022multi}. For TSBRs, some studies embed temporal signals into item representations to capture time-aware session preferences~\cite{zhou2021temporal, guo2025time}, while others incorporate temporal information into session graph structures to refine transition modeling~\cite{tang2022time,li2022spatiotemporal,wan2023spatio,chen2024combine}. Interval-aware methods further model temporal gaps between consecutive interactions to improve item-transition representation~\cite{wang2023interval,zuo2026time}. Moreover, multi-interest temporal methods use time information to distinguish heterogeneous short-term interests within the same session~\cite{wang2022time,shen2023temporal}. These designs improve the modeling of users' interests by exploiting temporal signals from different perspectives.

\section{Statistical Analysis}
\label{sec:stat}
In this section, we examine whether intervals should be treated as globally comparable across items. We first analyze the distributional mismatch between item-specific interval patterns and the overall interval distribution. We then further examine, under item-specific interval calibration, whether intervals located at different positions in their item-specific distributions contribute equally to recommendation performance.
\subsection{Intervals Are Not Globally Comparable}
\label{sec:item-distribution-disparity}

We first examine the extent to which temporal observations vary across items. For each item $i$, let $\mathcal{D}_i$ denote the set of its observed intervals and define the log-transformed observation set as $\mathcal{X}_i=\{\log(1+d):d\in\mathcal{D}_i\}$. We use $\mathcal{X}_i$ to construct the empirical item-specific interval distribution $\hat{F}_i(x)$ \cite{xu2008user,xu2011mining}. For the overall empirical interval distribution $\hat{F}_D(x)$, we aggregate all log-transformed intervals. We retain items with at least 20 temporal observations in the descriptive analysis. Our intuition is that if temporal signals were globally comparable across items, item-specific distributions $\hat{F}_i(x)$ should not depart substantially from $\hat{F}_D(x)$. We evaluate this assumption on three datasets.

\begin{figure*}[t]
  \centering
  \includegraphics[width=0.9\linewidth]{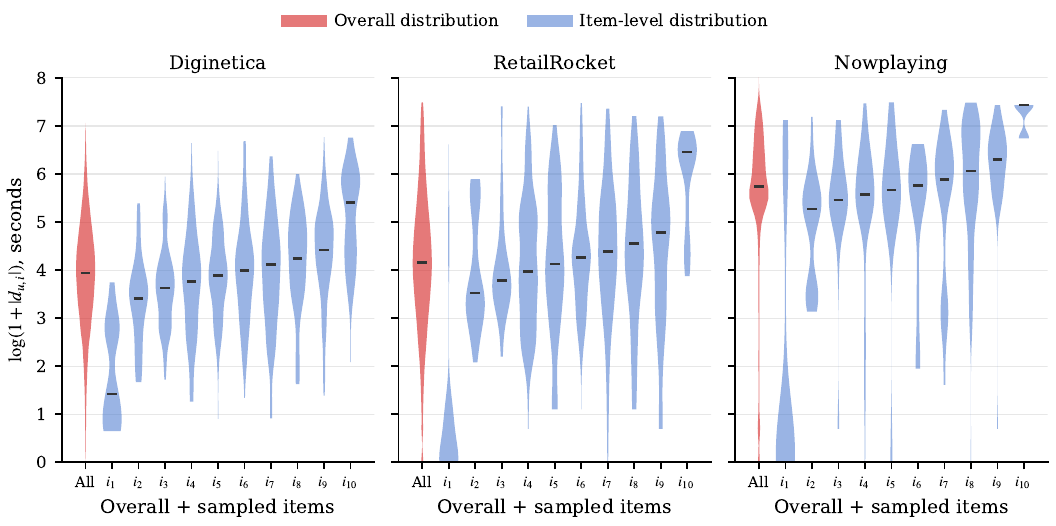}
  \caption{%
    Item-specific interval distribution disparity across datasets.
    The red violin denotes the overall interval distribution, and the blue violins denote ten representative item-specific distributions sampled across item-specific medians. The visible differences in location, dispersion, and tail behavior show that interval observations are not globally comparable across items.%
  }
  \label{fig:engagement-violin}
\end{figure*}

Figure~\ref{fig:engagement-violin} provides evidence of item-specific interval distribution disparity. The red violin denotes the overall interval distribution over eligible items, while the blue violins denote ten representative empirical item-specific distributions. 
The representative items are selected to span the low-to-high range of
item-specific median interval values. Across all three datasets, item-specific interval distributions visibly differ from the aggregate distribution and from one another. The differences involve central tendency, dispersion, and tail behavior. This descriptive pattern indicates that even a temporal value located around the median or high-density region of the overall interval distribution may fall into a lower or upper percentile range under an item-specific distribution. Hence, interval observations are not directly comparable across items without considering item-specific scales.

To assess whether the item-specific intervals distributional disparity is statistically supported, we conduct an aggregate-quantile chi-square homogeneity
test \cite{agresti2013categorical}. Our hypothesis is: \emph{ if items do not have a different interval distribution, then the intervals corresponding to an item should be distributed similarly to the overall interval distribution.} For the formal test, we sample 200 items in each dataset
and partition the aggregate distribution into \(K=10\) quantile bins
\(\{B_1,\ldots,B_K\}\). The null hypothesis is therefore
  \(H_0:\hat{F}_i(x)=\hat{F}_D(x),\forall i\), while the alternative is
  \(H_1:\exists i,\hat{F}_i(x)\neq\hat{F}_D(x)\). For item \(i\), the observed count in bin
  \(k\) is \(O_{ik}=\sum_{x\in\mathcal{X}_i}\mathbf{1}(x\in B_k)\), and the expected count
  under \(H_0\) is \(E_{ik}=n_ip_k\), where \(n_i=|\mathcal{X}_i|\) and \(p_k\approx 1/K\). The Pearson
statistic is computed as:
\begin{equation}
\chi^2=\sum_i\sum_k \frac{(O_{ik}-E_{ik})^2}{E_{ik}}.
\tag{I}
\end{equation}
\begin{table}[t]
\centering
\caption{Aggregate-quantile goodness-of-fit test between item-specific interval distributions and the overall interval distribution .}
\label{tab:item-aggregate-test}
\resizebox{\linewidth}{!}{%
\begin{tabular}{ccccc}
\toprule
Dataset & Values & $\chi^2$ & Cramer's $V$  & $p$-value \\
\midrule
Diginetica & 257,454 & 36,096.63  & 0.125 & $<0.01$ \\
RetailRocket & 199,550 & 39,972.79   & 0.149 & $<0.01$ \\
Nowplaying & 376,894 & 424,951.01 & 0.354 & $<0.01$ \\
\bottomrule
\end{tabular}%
}
\end{table}
As shown in Table~\ref{tab:item-aggregate-test}, the aggregate-quantile tests reject the null hypothesis. Because chi-square tests can become significant in large samples, we interpret the Pearson statistics together with Cramer's $V$ \cite{wasserstein2016asa}. The effect sizes indicate modest but non-trivial item-specific heterogeneity on Diginetica and RetailRocket, and a stronger discrepancy on Nowplaying. These results show that item-specific intervals cannot be treated as a single distribution.

\subsection{Impact of Item-specific Interval Disparity on Temporal Session Recommendation}

The distributional disparity analysis does not show the impact of interval distribution on recommendation performance directly. We therefore use a matched perturbation experiment to test whether item-specific interval typicality or atypicality corresponds to performance. 


We measure item-specific interval typicality using a two-standard-deviation range around the mean of item-specific intervals. We construct two matched groups of interactions. The outside group contains interactions whose intervals fall outside the two-standard-deviation range of their corresponding item-specific interval distributions. The inside group contains interactions that fall within the range, but are matched to the outside group by their percentile under the overall interval distribution. Thus, we could disturb the two groups with the same interval intervention to partition the impact of intervention magnitude on performance. Specifically, we perturb original intervals by one standard deviation, computed on the overall interval distribution. The perturbation is applied in a randomly sampled positive or negative direction; when the negative direction would produce an invalid negative interval, we resample the direction. For comparison, we use several GNN-based and Transformer-based TSBRs baselines: IGT, TMI-GNN, DT-GAT, and TE-GNN. Appendix~\ref{app:baselines} gives brief descriptions, and Table~\ref{tab:temporal_perturbation_results} reports the results.

\begin{table}[t] 
\centering
\caption{Impact of perturbing temporally typical and atypical interactions on Diginetica.}
\label{tab:temporal_perturbation_results}
\resizebox{\columnwidth}{!}{%
\begin{tabular}{llcccc}
\toprule
Model & Variant 
& Hit@5 & Hit@10 & MRR@5 & MRR@10 \\
\midrule

\multirow{5}{*}{IGT}
& Original            & 27.43 & 38.99 & 15.55 & 17.08 \\
& Inside              & 27.06 & 38.31 & 15.10 & 16.59 \\
& Inside change  & $-$0.37 & $-$0.68 & $-$0.45 & $-$0.49 \\
& Outside             & 27.61 & 38.92 & 15.70 & 17.21 \\
& Outside change &  0.18 & $-$0.07 &  0.15 &  0.13 \\
\midrule

\multirow{5}{*}{TMI\mbox{-}GNN}
& Original            & 25.95 & 37.23 & 14.56 & 16.07 \\
& Inside              & 25.55 & 36.79 & 14.10 & 15.58 \\
& Inside change  & $-$0.40 & $-$0.44 & $-$0.46 & $-$0.49 \\
& Outside             & 26.11 & 37.18 & 14.58 & 16.04 \\
& Outside change &  0.16 & $-$0.05 &  0.02 & $-$0.03 \\
\midrule

\multirow{5}{*}{DT\mbox{-}GAT}
& Original            & 28.19 & 39.19 & 16.36 & 17.83 \\
& Inside              & 27.38 & 38.66 & 15.81 & 17.31 \\
& Inside change  & $-$0.81 & $-$0.53 & $-$0.55 & $-$0.52 \\
& Outside             & 27.89 & 39.24 & 16.25 & 17.75 \\
& Outside change & $-$0.30 &  0.05 & $-$0.11 & $-$0.08 \\
\midrule

\multirow{5}{*}{TE\mbox{-}GNN}
& Original            & 28.08 & 39.18 & 15.58 & 17.10 \\
& Inside              & 27.49 & 38.53 & 15.28 & 16.74 \\
& Inside change  & $-$0.59 & $-$0.65 & $-$0.30 & $-$0.36 \\
& Outside             & 28.03 & 39.11 & 15.75 & 17.26 \\
& Outside change & $-$0.05 & $-$0.07 &  0.17 &  0.16 \\
\bottomrule
\end{tabular}%
}
\end{table}
Across all four models, perturbing the inside group degrades
performance on both HR and MRR metrics, with particularly larger
drops on MRR@5. This indicates that the interval within the item-specific typical range provides more stable evidence for modeling short-term intent. Perturbing the outside group does not produce a stable performance decline; in some cases, it even leads to marginal improvements. This suggests that atypical intervals in the two tails are not consistently informative for TSBRs and may sometimes act as noisy interval signals. The experiment serves as controlled diagnostic evidence that item-specific interval typicality affects the stability of temporal signals during training.

Above analyzes give two takeaways. First, temporal values are not globally comparable across items: a value that appears typical under the dataset-level distribution may be atypical under the corresponding item-specific distribution. Second, item-specific interval typicality in its distribution is associated with higher recommendation performance, since perturbing temporally typical interactions causes performance drops compared with perturbing atypical interactions. 

\section{Methodology}
\label{sec:method}

\begin{figure*}[t] 
\centering 
\includegraphics[width=0.9\linewidth]{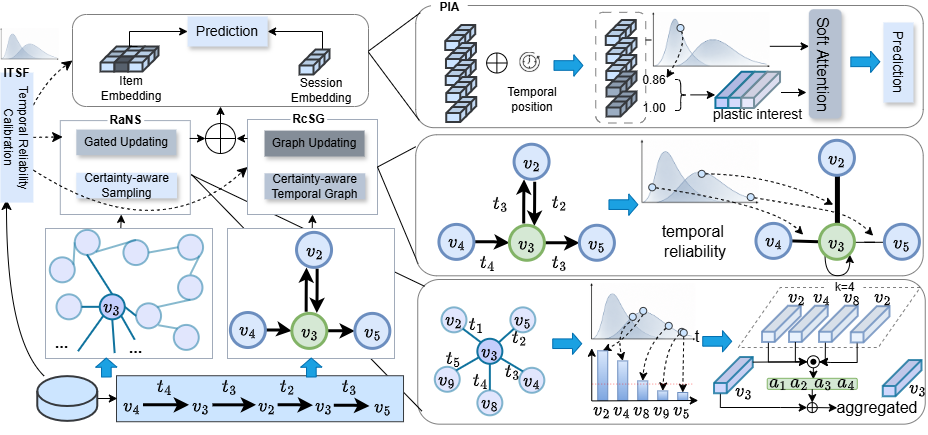} 
\caption{ Overview of the proposed TRUST framework. Item-calibrated temporal reliability is estimated from empirical item-specific interval distributions and used to sample reliable global neighbors, weight session-graph transitions, and capture reliable interests during session representation generation. } 
\label{fig:framework} 
\end{figure*}

The analysis in \S\ref{sec:item-distribution-disparity} suggests that item-specific interval heterogeneity can limit interval signals when a model treats them through absolute value. TRUST addresses this issue through three stages, as illustrated in \Cref{fig:framework}: 
(i) calibrating each available item-interval observation into a temporal reliability score using the item's empirical interval reference; 
(ii) using this shared reliability signal to guide item-specific representation learning from both reliability-aware global neighbor sampling and reliability-calibrated session graph encoding; 
and (iii) generating the final session representation through reliability-constrained plastic interest aggregation before next-item prediction. 

\noindent  \textbf{\textit{Problem Formulation}}

We begin by defining the notations used. Let $\Iset = \{1, \ldots, |\Iset|\}$ be the items set.
A prediction instance consists of an observed item sequence $\mathcal{S}=[v_1,\ldots,v_n]$, where $v_i \in \Iset$ is the item at position $i$.
For each observed transition from $v_i$ to $v_{i+1}$, let $d_i$ denote the interval observation on item $v_i$. 
Our goal is to predict the next item $v_{n+1}$ given $v_1, \ldots, v_n$ and corresponding intervals. 


\subsection{Shared Item-Calibrated Temporal Scoring Function}
\label{sec:temporal-reliability}

We first define the shared Item-Calibrated Temporal Scoring Function (ITSF) used throughout TRUST. Building on \Cref{sec:item-distribution-disparity}, our goal is to measure how much an observed interval $d_i$ deviates from the interval distribution of other users on the same item $v_i$. In our framework, such deviation is treated as evidence that the observed behavior is less consistent with the item-specific temporal pattern, and thus should receive a lower temporal reliability score. Moreover, this evidence should be weighted by the number of observations available for the item \cite{kuleshov2018accurate,han2022addressing}: when an item has only a few intervals observed, the item-level temporal distribution is uncertain, and the model should be conservative in penalizing deviations. Therefore, we define the Item-Calibrated Temporal Scoring Function, which computes the temporal reliability score $q_i(d_i)$ as:
\begin{equation}
\begin{aligned}
q_i(d_i)
&=
\exp\left[
-\frac{1}{2}
\left(
\frac{|z_i(d_i)|}
{1+\frac{1}{\sqrt{n_i+\epsilon_q}}}
\right)^\gamma
\right],
\end{aligned}
  \label{eq:uncertainty-temporal-reliability}
\end{equation}
where $\gamma>0$ controls how sharply reliability decays as the effective temporal deviation increases, $\epsilon_q$ is a small constant for numerical stability, and 
$n_i$ is the number of observed intervals for the corresponding item $i$. 
A smaller $n_i$ reflects greater uncertainty and therefore relaxes the penalty 
associated with the temporal deviation $|z_i(d_i)|$. It measures the empirical quantile position of interval $d_i$ in $v_i$'s item-specific intervals distribution, which $z_i(d_i)$ is defined as: 
\begin{equation}
z_i(d_i)
=
\Phi^{-1}
\left(
\operatorname{clip}
\left(
\hat{F}_i(\log(1+d_i)),
\rho,
1-\rho
\right)
\right),
\end{equation}
where $\Phi^{-1}(\cdot)$ is the inverse standard normal cumulative distribution function, and $\rho$ is a small clipping constant used to avoid infinite values at the distribution tails.  



\subsection{Reliability-Guided Structural-Temporal Encoding}
\label{sec:dual-channel}

Using the ITSF from \S\ref{sec:temporal-reliability}, we next describe how TRUST incorporates this shared temporal evidence into item-level representation learning. 
As shown in Fig.~\ref{fig:framework}, encoders consist of two reliability-guided message sources: a global source obtained through reliability-aware neighbor sampling (RaNS), and a local source obtained from a reliability-calibrated session graph (RcSG). 

\subsubsection{Reliability-aware Global Graph Encoding (RaNS)}
The global co-occurrence graph provides collaborative evidence beyond an individual session, but its neighbors are formed from all observed consecutive interactions and may therefore include relations supported mainly by temporally unreliable behavior. Directly sampling from this graph can propagate such noisy collaborative signals into item representations. The global branch addresses this challenge by using RaNS to prioritize neighbors supported by reliable intervals and then applying graph convolution to encode the resulting collaborative evidence.

To implement RaNS, we first construct an undirected global item co-occurrence graph 
$\mathcal{G}_g=(\mathcal{V},\mathcal{E}_g)$ only from training sessions, where an edge 
$\{i,j\}$ indicates that items $i$ and $j$ appear in a pair of consecutive interactions in at least one session. All interactions from item $i$ to item $j$ comprise a set $\mathcal{O}_{i\rightarrow j}$ 

After constructing the global graph, we further sample item neighbors from it.  Because the sampled quality directly shapes the information propagated during graph representation learning, , instead of randomly sampling neighboring items \cite{deng2022g,ou2025ls}, we choose the neighboring items that have the strongest reliable collaborative signal. We quantify this signal using the edge weight. To filter out noise connections, we only retain edges whose aggregated reliability weight exceeds the threshold $p$, and sample $w$ neighboring items to form the neighboring sampling set $\tilde{\mathcal{N}}_g(i)$:

\begin{equation}
\begin{gathered}
  \tilde{\mathcal{N}}_g(i)
  =
  \operatorname{Sample}_w
  \left(
  \{j \in \mathcal{N}_g(i) \mid \omega_{\{i,j\}}>p\}
  \right),\\
  \omega_{\{i,j\}}
  =
  \sum_{o\in\mathcal{O}_{i\rightarrow j}}
  q_i(d_o)
  +
  \sum_{o\in\mathcal{O}_{j\rightarrow i}}
  q_j(d_o),
\end{gathered}
\label{eq:retained-global-neighbor}
\end{equation}
where edge weights $\omega_{\{i,j\}}$ aggregate both the frequency of item co-occurrences and the temporal reliability of each observed interval, and $d_o$ denotes the interval in observation $o$. $\mathcal{N}_g(i)$ is the whole set of neighboring items in the graph. If $\tilde{\mathcal{N}}_g(i)$ is empty, we retain the self-loop instead of sampling. Otherwise, $\operatorname{Sample}(\cdot)$ is a probability-based sampling function with normalized weight $P(j\mid i)=\omega_{\{i,j\}}/\sum_{j'\in\mathcal{C}_g(i)}\omega_{\{i,j'\}}$, which is defined as:
\begin{itemize}
\item If $|\mathcal{C}_g(i)| \geq w$, we sample $w$ neighbors without replacement according to $P(j\mid i)$. 
\item If $|\mathcal{C}_g(i)|< w$, all retained neighbors are first included, and the remaining $w-|\mathcal{C}_g(i)|$ positions are filled by sampling with replacement according to $P(j\mid i)$.
\end{itemize}

After obtaining $\tilde{\mathcal{N}}_g(i)$, we apply a gated graph convolution to aggregate global collaborative evidence. 
Let $\mathbf{h}_i$ denote the representation of item $i$, initialized by its learnable item embedding. 
We finally compute global graph convolution as $ \mathbf{m}_{i}^{(g)} $:
\begin{equation}
\begin{aligned}
  \mathbf{m}_{i}^{(g)}
  =
  \sum_{j\in \tilde{\mathcal{N}}_g(i)}
  a_{ij}^{(g)}
  \mathbf{W}_g\mathbf{h}_{j},
  \\
    a_{ij}^{(g)}
  =
  \sigma\left(
  \mathbf{w}_g^\top
  [\mathbf{h}_i;\mathbf{h}_j]
  + b_g
  \right),
\end{aligned}
\label{eq:global-message}
\end{equation}
where $\mathbf{W}_g$ is a learnable transformation matrix, $\mathbf{w}_g$ and $b_g$ are learnable parameters.

\subsubsection{Reliability-calibrated Session Graph (RcSG)}
As discussed in \Cref{sec:stat}, although TSBR graph encoders utilize absolute intervals, they cannot account for distributional differences in intervals across items. RcSG addresses this by incorporating ITSF scores into directed session-graph message passing, so that local transitions are encoded according to their item-calibrated temporal reliability.

To do so, it constructs a directed session graph and uses the ITSF score as convolutional edge weights. 
We fuse temporal reliability with session graph learning through a temporal reliability-aware attention mechanism. Each item $i$'s representation $\mathbf{m}_{i}^{(s)}$ is then computed as:
\begin{equation}
\label{eq:session-local-attn}
\begin{gathered}
\mathbf{m}_{i}^{(s)}
=
\sum_{j\in\mathcal{N}_s(i)}
\beta_{ij}^{(s)}
\mathbf{h}_j,
\\
\beta_{ij}^{(s)}
=
\operatorname{softmax}_{j\in\mathcal{N}_s(i)}
\left(e_{ij}^{(s)}\right),
\\
e_{ij}^{(s)}
=
\operatorname{LeakyReLU}
\Bigl(
\mathbf{W}_1^{s}
\bigl(
\mathbf{h}_i \odot \mathbf{h}_j
\bigr)
+
\mathbf{W}_2^{s}
\mathcal{T}(q_i(d_i))
\Bigr).
\end{gathered}
\end{equation}
where, $\mathcal{N}_s(i)$ denotes the outgoing neighboring items of $i$ in the directed session graph, and $\odot$ denotes element-wise multiplication. $\mathcal{T}(\cdot)$ is a linear projection that maps the scalar reliability score into a $d$-dimensional temporal feature, while $\mathbf{W}_1^{s}$ and $\mathbf{W}_2^{s}$ project their inputs to scalar edge scores before the LeakyReLU activation.

\subsection{Plastic Interest Aggregation (PIA)}
\label{sec:plastic-agg}

A remaining challenge lies in how to effectively fuse the item representations learned by the two encoders into a unified session representation. This is non-trivial because existing methods commonly treat the last item as the short-term interest signal for matching other items to generate a session representation, although the last item alone may be noisy and insufficient. To address it, we propose Plastic Interest Aggregation, which models reliable short-term interests at multiple depths.

We first fuse two encoders' outputs with temporal position information to get the fused item $i$'s representation $\bar{\mathbf{h}}_{i}$: 
\begin{equation}
\begin{aligned}
  \bar{\mathbf{h}}_{i}
  &=
  \mathbf{m}_{i}^{(g)}
  +
  \mathbf{m}_{i}^{(s)}
  +
  \mathbf{p}_{\ell_i}^{(\mathrm{rev})},
  \quad
  \ell_i = n-i+1,
\end{aligned}
\label{eq:reverse-position-fusion}
\end{equation}
where $\ell_i$ is the reverse position index, and $\mathbf{p}_{\ell_i}^{(\mathrm{rev})}\in\mathbb{R}^{d}$ is a learnable reverse positional embedding.

PIA then builds reliable interest queries at multiple depths. We assign a position-based reliability score to each position. To prevent data leakage, we assign the last position score as 1, the position-based reliability score $\tilde{r}_i$ is:
\begin{equation}
  \tilde{r}_i =
  \begin{cases}
    q_{i}(d_i), & 1\le i<n,\\
    1, & i=n,
  \end{cases}
  \label{eq:reliability-coordinate-mass}
\end{equation}
where $d_i$ is the observed interval on the item at position $i$.
For each depth-$b$ view, PIA allocates a depth of $b \in \mathcal{B}$ backward from the last item and assigns the position weight $\xi_i^{(b)}$ as follows:
\begin{equation}
  \xi_i^{(b)}
  =
  \max\!\left(
  0,
  \min\!\left(
  \tilde{r}_i,
  b-\sum_{k=i+1}^{n}\xi_k^{(b)}
  \right)
  \right),
  \label{eq:budget-allocation}
\end{equation}
where the weights are computed recursively for $i=n,n-1,\ldots,1$. Small depths focus on the most recent reliable items; larger depths extend toward earlier items. 
After normalizing $\xi_i^{(b)}$ as $\bar{\xi}_i^{(b)}=\xi_i^{(b)}/(\sum_{j=1}^{n}\xi_j^{(b)}+\epsilon_{\mathrm{norm}})$, where $\epsilon_{\mathrm{norm}}$ is a small constant for numerical stability, the depth-specific plastic query $\mathbf{q}_p^{(b)}$ is:
\begin{equation}
  \mathbf{q}_p^{(b)}
  =
  \sum_{i=1}^{n}
  \bar{\xi}_i^{(b)}
  \bar{\mathbf{h}}_{i}.
  \label{eq:plastic-query}
\end{equation}

For each query $\mathbf{q}_p^{(b)}$ with depth $b$, we apply additive attention over all items in the session to obtain the various view of session representation $\mathbf{s}^{(b)}$:
\begin{equation}
\begin{gathered}
\mathbf{s}^{(b)}
=
\sum_{i=1}^{n}
\beta_i^{(b)}
\bar{\mathbf{h}}_{i},
\\
\beta_i^{(b)}
=
\frac{
\exp(a_i^{(b)})
}{
\sum_{j=1}^{n}
\exp(a_j^{(b)})
},
\\
a_i^{(b)}
=
\mathbf{w}_a^\top
\sigma \!\left(
\mathbf{W}_a \bar{\mathbf{h}}_{i}
+
\mathbf{W}_q \mathbf{q}_p^{(b)}
+
\mathbf{b}_a
\right),
\end{gathered}
\label{eq:reliability-constrained-attn}
\end{equation}
where $\sigma$ is the Sigmoid activation, $\mathbf{W}_a,\mathbf{W}_q\in\mathbb{R}^{d\times d}$ and $\mathbf{w}_a\in\mathbb{R}^{d}$ are learnable parameters. 
The final session representation $\mathbf{s}$ is obtained by averaging the available views:
\begin{equation}
  \mathbf{s}
  =
  \frac{1}{|\mathcal{B}|}
  \sum_{b\in\mathcal{B}}
  \mathbf{s}^{(b)}.
  \label{eq:session-rep}
\end{equation}
To this end, we complete the modelling of our proposed framework. The complete procedure is summarized in \Cref{Alg:trust}.

\textbf{\textit{Training Objective:}}


The training objective is to minimize the cross-entropy loss $\mathcal{L} $ over the full item softmax:
\begin{equation}
\begin{aligned}
  \mathcal{L}
  &= -\log \frac{\exp(\hat{y}_{v^*})}
  {\sum_{j \in \Iset} \exp(\hat{y}_j)}, \\
  \hat{y}_j
  &= \mathbf{s}^\top \mathbf{h}_j,
  \quad \forall\, j \in \Iset .
\end{aligned}
\label{eq:loss}
\end{equation}
where $v^*=v_{n+1}$ is the ground-truth next item, $\mathbf{h}_j \in \mathbb{R}^d$ is the shared embedding of item $j$, and $\Iset$ is the set of all items.  

\begin{algorithm}[t]
\caption{TRUST: Temporal Reliability Calibration Framework}
\label{Alg:trust}
\begin{algorithmic}[1]
    \STATE \textbf{Input:} Session $\mathcal{S}$, item set $\mathcal{I}$, training sessions $\mathcal{D}$.
    \STATE \textbf{Output:} Ranked items for next-item prediction.
    
    \STATE Estimate item-specific interval references and compute ITSF reliability scores.
    \hfill [Eq.~\ref{eq:uncertainty-temporal-reliability}]
    
    \STATE Construct reliability-aware global graph with $w$ neighboring sampling size and compute global messages.
    \hfill [Eqs.~\ref{eq:retained-global-neighbor}--\ref{eq:global-message}]
    
    \STATE Compute reliability-calibrated session messages and fuse item representations.
    \hfill [Eqs.~\ref{eq:session-local-attn}, \ref{eq:reverse-position-fusion}]
    
    \STATE Build leakage-free reliability weights for session-level interest aggregation.
    \hfill [Eq.~\ref{eq:reliability-coordinate-mass}]
    
    \STATE Generate multi-granularity plastic interest queries.
    \hfill [Eqs.~\ref{eq:budget-allocation}--\ref{eq:plastic-query}]
    
    \STATE Compute granularity-specific attention and fuse session representations. \hfill\mbox{[Eqs.~\ref{eq:reliability-constrained-attn}--\ref{eq:session-rep}]}
    
    \STATE Compute scores, optimize objective, and return ranked items.
    \hfill [Eq.~\ref{eq:loss}]
\end{algorithmic}
\end{algorithm}

\section{Experiments}
\label{sec:experiments}

\subsection{Experimental Setup}

\subsubsection{Datasets}
We conduct experiments on three widely used public datasets: {Diginetica}, {RetailRocket}, and {Nowplaying}. Diginetica is derived from the CIKM Cup 2016 competition \cite{zhao2015commerce} and contains user click sequences on product pages, each annotated with the time spent. RetailRocket covers product-view, add-to-cart, and purchase events; following standard protocol we retain only view events \cite{wang2021survey}. Nowplaying is a music-domain dataset constructed from listening list, where each session records successive song interactions \cite{ludewig2018evaluation}. For all three datasets, we follow the standard preprocessing procedure~\cite{pan2020star,yu2023causality,zhang2025survey}: sessions of length one are discarded, items appearing fewer than five times are removed, and the last item in each session is held out as the prediction target. Table~\ref{tab:datasets} summarizes the statistics of datasets. The column Mean/Med. gives the ratio of the mean to the median of intervals.

\begin{table}[t]
\centering
\caption{Dataset statistics of the three datasets}
\label{tab:datasets}
\resizebox{\columnwidth}{!}{%
\begin{tabular}{lrrrrr}
\toprule
Dataset      & \#Items & Train sess. & Test sess. & Avg.\ len & Mean/Med. \\
\midrule
Diginetica   & 43,097  & 719,470 & 60,858 & 5.12 & 1.71 \\
RetailRocket & 50,018  & 721,266 & 61,969 & 4.35 & 2.56 \\
Nowplaying   & 60,416  & 825,304 & 89,824 & 5.55 & 1.45 \\
\bottomrule
\end{tabular}}
\end{table}

\subsubsection{Evaluation Metrics}
We adopt Hit Rate (HR@$K$) and Mean Reciprocal Rank (MRR@$K$) at $K\in\{5,10\}$. HR@$K$ measures whether the ground-truth next item appears in the top-$K$ predicted list; MRR@$K$ suggests whether a high rank for the ground-truth item. All metrics are reported as percentages.

\subsubsection{Baselines}
We compare TRUST against two groups of baselines. The \emph{non-temporal} SBR group includes GRU4Rec~\cite{hidasi2015session}, NARM~\cite{li2017neural}, SR-GNN~\cite{wu2019session}, FGNN~\cite{qiu2019rethinking}, GCE-GNN~\cite{wang2020global}, $S^2$-DHCN~\cite{xia2021selfhypergraph}, MSGAT~\cite{qiao2023bi}, and DMI-GNN \cite{lv2025dynamic}. The TSBRs group includes STAN~\cite{garg2019sequence}, TASRec~\cite{zhou2021temporal}, TE-GNN~\cite{tang2022time}, IGT~\cite{wang2023interval}, TMI-GNN~\cite{shen2023temporal}, and DT-GAT~\cite{guo2026dual}. The baselines description lies in Appendix~\ref{app:baselines}.
\begin{table*}[t]
\centering
\caption{Overall recommendation performance (\%). Best result per column is \textbf{bold}; second-best is \underline{underlined}. `* indicates the difference between TRUST and the second-best model is statistically significant ($p$-value $<$ 0.01 by a paired $t$-test based on 5 runs of the best performing model and the second-best performing model, respectively)}
\label{tab:main-results}
\resizebox{\textwidth}{!}{%
\begin{tabular}{lcccccccccccc}
\toprule
& \multicolumn{4}{c}{\textbf{Diginetica}}
& \multicolumn{4}{c}{\textbf{RetailRocket}}
& \multicolumn{4}{c}{\textbf{Nowplaying}} \\
\cmidrule(lr){2-5}\cmidrule(lr){6-9}\cmidrule(lr){10-13}
Method & HR@5 & HR@10 & MRR@5 & MRR@10 & HR@5 & HR@10 & MRR@5 & MRR@10 & HR@5 & HR@10 & MRR@5 & MRR@10 \\
\midrule
GRU4Rec    & 11.93 & 18.53 & 6.46  & 7.49  & 43.16 & 47.71 & 33.94 & 35.20 & 7.42  & 11.18 & 4.30  & 4.88  \\
NARM       & 25.95 & 35.55 & 14.07 & 15.38 & 47.84 & 55.47 & 34.20 & 35.08 & 8.15  & 12.34 & 4.86  & 5.41  \\
SR-GNN     & 27.15 & 38.65 & 15.29 & 16.82 & 48.66 & 56.04 & 36.25 & 37.24 & 9.31  & 13.67 & 5.52  & 6.10  \\
FGNN       & 24.39 & 35.90 & 13.22 & 14.64 & 47.35 & 55.29 & 35.24 & 35.36 & 8.63  & 12.89 & 5.07  & 5.52  \\
$S^2$-DHCN & \underline{28.35} & 39.87 & 15.68 & 17.53 & 48.94 & 56.94 & 36.15 & 36.92 & 9.74  & 14.05 & 5.76  & 6.35  \\
MSGAT      & 24.74 & 35.58 & 13.99 & 15.47 & 47.92 & 56.00 & 35.72 & 36.71 & 8.92  & 13.21 & 5.31  & 5.87  \\
DMI-GNN    & 28.08 & \underline{40.16} & 16.27 & \underline{17.91} & \underline{51.26} & \underline{58.05} & \underline{38.20} & \underline{39.17} & \underline{12.34} & \underline{17.09} & \underline{7.67} & \underline{8.01} \\
\midrule
STAN       & 25.71 & 38.70 & 15.64 & 17.12 & 49.40 & 56.28 & 37.75 & 38.68 & 9.03  & 13.04 & 5.34  & 5.88  \\
TASRec     & 27.39 & 39.10 & 15.31 & 17.35 & 51.03 & 57.86 & 38.16 & 39.13 & 10.45 & 15.22 & 6.18  & 6.79  \\
TE-GNN     & 28.08 & 39.18 & 15.58 & 17.10 & 47.40 & 54.85 & 34.98 & 35.98 & 11.10 & 15.91 & 6.60  & 7.23  \\
IGT        & 27.43 & 38.99 & 15.55 & 17.08 & 48.42 & 55.71 & 36.20 & 36.96 & 10.82 & 15.47 & 6.43  & 7.03  \\
TMI-GNN    & 25.95 & 37.23 & 14.56 & 16.07 & 50.58 & 57.87 & 37.45 & 38.44 & 9.77  & 14.32 & 5.80  & 6.41  \\
DT-GAT     & 28.19 & 39.19 & \underline{16.36} & 17.83 & 49.70 & 56.24 & 36.48 & 37.36 & 9.70  & 14.77 & 6.03  & 6.77  \\
\midrule
\textbf{TRUST*} & \textbf{30.07*} & \textbf{41.79*} & \textbf{17.27*} & \textbf{18.82*} & \textbf{52.30*} & \textbf{59.87*} & \textbf{39.10*} & \textbf{40.22*} & \textbf{12.68*} & \textbf{17.35} & \textbf{7.79*} & \textbf{8.40*} \\
\midrule
Improve & $+6.07\%$ & $+4.06\%$ & $+5.56\%$ & $+5.08\%$
        & $+2.03\%$ & $+3.14\%$ & $+2.36\%$ & $+2.68\%$
        & $+2.76\%$ & $+1.52\%$ & $+1.56\%$ & $+4.87\%$ \\
\bottomrule
\end{tabular}}
\end{table*}
\subsubsection{Implementation Details}
\label{sec:impl-details}
All parameters are initialized with Xavier-uniform initialization and optimized with Adam at an initial learning rate of $10^{-3}$. Both item-embedding dimension and batch size is set as 100 to align with baselines. We apply early stopping based on HR@10, with a patience of 3 epochs. The ITSF is first estimated from the training dataset before training. The clipping constant $\rho$ is set at 0.01. The multi-granularity plastic interest in PIA is set as $\mathcal{B}=\{1,2,3\}$. 


\subsection{Research Questions}
The experiments evaluate four aspects: overall recommendation accuracy, plug-in (ITSF) applicability to existing TSBRs, the contribution of each reliability-guided use of the signal (ablation), and computational practicality. Five questions are as follows: RQ1 evaluates whether TRUST improves next-item recommendation over non-temporal and temporal baselines. RQ2 tests whether the calibrated reliability signal is useful beyond TRUST as a plug-in replacement for raw interval. RQ3 isolates whether the three uses of the reliability signal, global sampling, session-graph transition weighting, and interest aggregation, each contribute to performance. RQ4 analyzes the sensitivity of TRUST to the main hyperparameters. RQ5 examines algorithm efficiency.

\subsection{RQ1: Overall Performance Comparison}

Table~\ref{tab:main-results} reports the recommendation performance on Diginetica, RetailRocket, and Nowplaying. TRUST achieves the best result on almost all metrics and shows statistically significant gains over the strongest competing baselines in most cases. The magnitude of improvement varies across datasets, while gains remain consistent. 

TBSRs do not always outperform strong non-temporal SBRs. For example, several temporal methods perform below DMI-GNN on Diginetica and Nowplaying. On the one hand, DMI-GNN is a recent state-of-the-art model that utilizes a dynamic multi-interest mechanism. Its performance also suggests that raw temporal intervals alone do not guarantee better recommendation accuracy. TRUST addresses this issue by calibrating temporal information at the item level before using it as evidence of user interest.

\subsection{RQ2: Plug-In Applicability with ITSF}
\label{sec:rq2}

We also test whether the item-calibrated temporal signal helps models beyond the full TRUST architecture. Specifically, we integrate ITSF into four temporal baselines: IGT, TE-GNN, TMI-GNN, and DT-GAT. We utilize ITSF to replace the original interval used in these methods, while the session representation generation, training objective, optimizer, and other implementation details remain unchanged. We evaluate the resulting variants under the same training protocol as \S\ref{sec:impl-details}.

Table~\ref{tab:plugin} compares the original temporal baselines with their ITSF-based variants. Replacing the original interval with ITSF improves HR@10 and MRR@10 for all four baselines. The gains vary across backbones and datasets, but the results show that ITSF could serve as a model-agnostic module for TSBRs.

\begin{table}[t]
\centering
\caption{Performance of temporal SBR backbones with and without ITSF. ``Cal.'' denotes the variant integrated with ITSF.}
\label{tab:plugin}
\small
\setlength{\tabcolsep}{4pt}
\resizebox{\columnwidth}{!}{%
\begin{tabular}{lcccc}
\toprule
& \multicolumn{2}{c}{\textbf{Diginetica}}
& \multicolumn{2}{c}{\textbf{Nowplaying}} \\
\cmidrule(lr){2-3}\cmidrule(lr){4-5}
Method
& HR@10 & MRR@10
& HR@10 & MRR@10 \\
\midrule
IGT         & 38.99 & 17.08 & 15.47 & 7.03 \\
Cal.        & 39.24 & 17.38 & 15.94 & 7.47 \\
Gain   & +0.64\% & +1.76\% & +3.04\% & +6.26\% \\
\midrule
TE-GNN      & 39.18 & 17.10 & 15.91 & 7.23 \\
Cal.        & 39.53 & 17.46 & 16.37 & 7.28 \\
Gain   & +0.89\% & +2.11\% & +2.89\% & +0.69\% \\
\midrule
TMI-GNN     & 37.23 & 16.07 & 14.32 & 6.41 \\
Cal.        & 37.84 & 16.45 & 14.97 & 6.62 \\
Gain   & +1.64\% & +2.36\% & +4.54\% & +3.28\% \\
\midrule
DT-GAT      & 39.19 & 17.83 & 14.77 & 6.77 \\
Cal.        & 40.13 & 18.34 & 15.16 & 7.18 \\
Gain   & +2.40\% & +2.86\% & +2.64\% & +6.06\% \\
\bottomrule
\end{tabular}}
\end{table}


\subsection{RQ3: Ablation Study}
\label{sec:rq3}

In this experiment, we verify the contribution of each TRUST component, especially the item-calibrated temporal signal in each module, by evaluating the following variants:

\begin{itemize}[leftmargin=1.5em,itemsep=0pt,topsep=2pt]
\item \textbf{w/o RaNS\_cal.} Replace reliability-aware neighbor sampling with uniform co-occurrence-frequency sampling, removing the reliability-aware neighbor sampling in Eq.~\eqref{eq:retained-global-neighbor}.
\item \textbf{w/o RcSG\_cal.} Remove the temporal reliability from the session-graph learning in Eq.~\eqref{eq:session-local-attn}, reducing it to standard graph attention.
\item \textbf{w UniPIA.} Keep the PIA aggregation structure but set all reliability weights in PIA to one, so the backward multi-granularity plastic interest aggregation remains while reliability-weighted allocation is removed.
\item \textbf{w LastQ.} Replace PIA with a standard short-term interest-based session generation, which uses the last item as the short-term query.
\item \textbf{w/o ITSF.} Replace the item-calibrated temporal scoring function with the original absolute intervals, so that scoring is no longer estimated relative to each item's empirical interval distribution.
\end{itemize}

Table~\ref{tab:ablation} reports the ablation results. Each variant performs worse than the full model, but the magnitude of the drop varies by module. Performance decreases when RaNS is removed, suggesting that reliability-aware neighbor sampling helps prevent noisy global co-occurrence relations in global graph learning. Removing RcSG also reduces performance, especially on Nowplaying, where local transition reliability appears more important.

The two PIA-based variants give a direct view of the session generation step. UniPIA remains competitive because it maintains the multi-granularity plastic interest aggregation structure, but it falls slightly below full TRUST when reliability-weighted allocation is removed. LastQ is also close to the full model on Diginetica, yet it is weaker on both datasets, especially on Nowplaying MRR@10. These results suggest that the PIA structure matters, and the calibrated reliability weights still provide useful information when deciding how much recent interactions should contribute.

Removing ITSF results in a larger decline than removing either RaNS or RcSG alone, because ITSF provides the calibrated reliability signal used by the downstream reliability-guided modules. Once ITSF is replaced by absolute intervals, the model can no longer distinguish whether an observed interval is typical for the specific item under consideration.

\begin{table}[t]
\centering
\caption{Ablation results on Diginetica and Nowplaying (\%). Best result per column is \textbf{bold}.}
\label{tab:ablation}
\small
\setlength{\tabcolsep}{4pt}
\begin{tabular}{lcccc}
\toprule
& \multicolumn{2}{c}{\textbf{Diginetica}}
& \multicolumn{2}{c}{\textbf{Nowplaying}} \\
\cmidrule(lr){2-3}\cmidrule(lr){4-5}
Variant
& HR@10 & MRR@10
& HR@10 & MRR@10 \\
\midrule

w/o RaNS\_cal
& 41.36 & 18.33
& 17.27 & 8.16 \\
w/o RcSG\_cal
& 41.44 & 18.23
& 16.99 & 7.97 \\
w UniPIA
& 41.29 & 18.47
& 16.97 & 8.14 \\
w LastQ
& 41.40 & 18.50
& 17.18 & 8.09 \\
\midrule

w/o ITSF
& 39.87 & 17.54
& 16.42 & 7.81 \\
\midrule

Full TRUST
& \textbf{41.79} & \textbf{18.82}
& \textbf{17.35} & \textbf{8.40} \\
\bottomrule
\end{tabular}
\end{table}

\subsection{RQ4: Hyperparameter Sensitivity}
\label{sec:rq4}

We examine two hyperparameters: the neighbor sampling size $w$ on global graph and the temporal reliability decay exponent $\gamma$ on ITSF. Figure~\ref{fig:hyper-window} and Figure~\ref{fig:hyper-gamma} report the results.

\subsubsection{Effect of neighbor sampling size $w$.}
\begin{figure*}[th]
\centering
\includegraphics[width=0.25\linewidth]{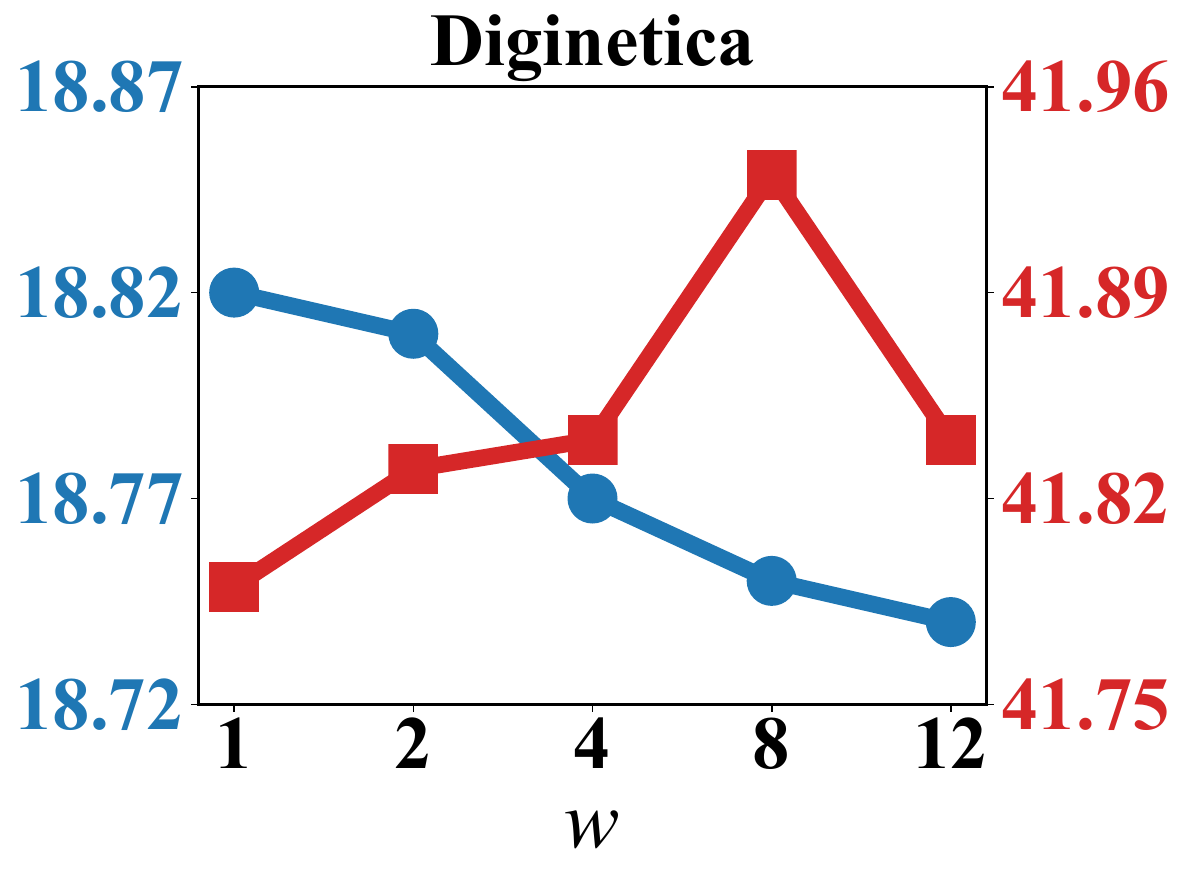}
\includegraphics[width=0.25\linewidth]{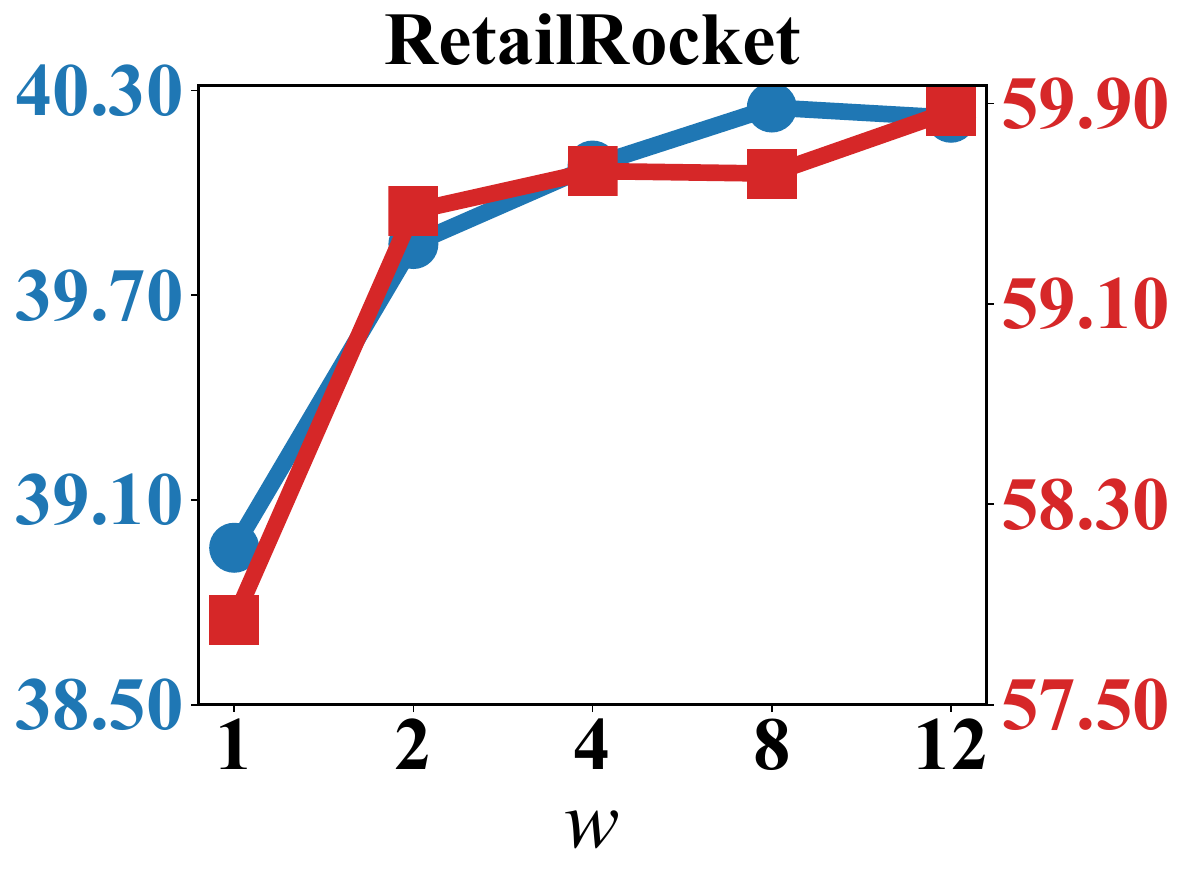}
\includegraphics[width=0.25\linewidth]{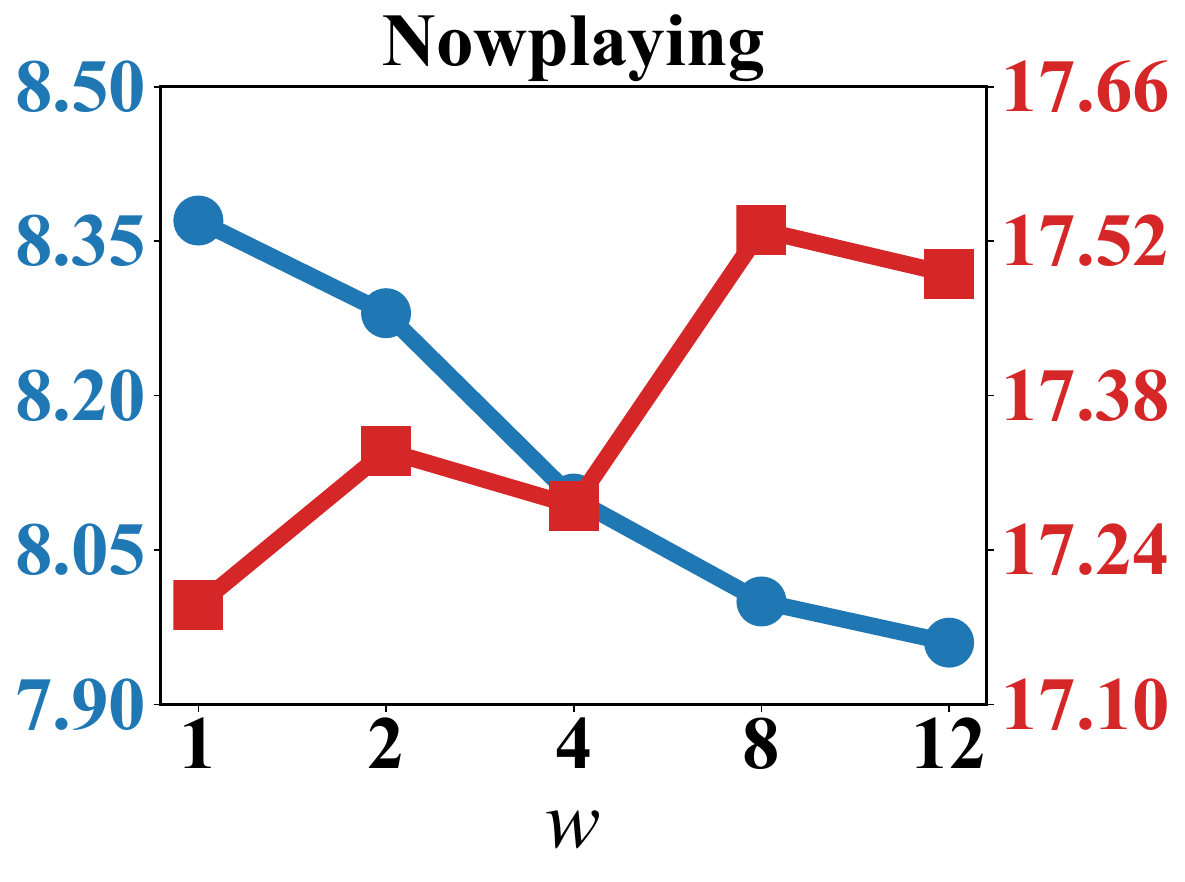}
\caption{Hyperparameter sensitivity of TRUST with respect to neighbor sampling size $w$. HR@10 and MRR@10 are reported on Diginetica, RetailRocket, and Nowplaying.}
\label{fig:hyper-window}
\end{figure*}

The neighbor sampling size $w$ controls the number of neighboring items used in RaNS for graph learning. As shown in Figure~\ref{fig:hyper-window}, the effect of $w$ is dataset-dependent but generally moderate. On Diginetica, HR@10 increases slightly as $w$ grows, reaching its maximum at $w=8$, whereas MRR@10 remains stable and is marginally higher for smaller neighbor sampling sizes. On Nowplaying, HR@10 also benefits from a larger sampling size and peaks at $w=8$, while MRR@10 gradually decreases as $w$ increases, suggesting that overly broad sampling may weaken top-rank precision in the music-domain setting. RetailRocket shows a different pattern: both HR@10 and MRR@10 improve substantially as $w$ increases, with the best ranking quality around $w=8$ and the highest HR@10 at $w=12$. In the main experiments, we choose $w$ based on validation performance, resulting in $w=1$ for Diginetica, $w=12$ for RetailRocket, and $w=2$ for Nowplaying.

\subsubsection{Effect of decay exponent $\gamma$.}
\begin{figure*}[t]
\centering
\includegraphics[width=0.25\linewidth]{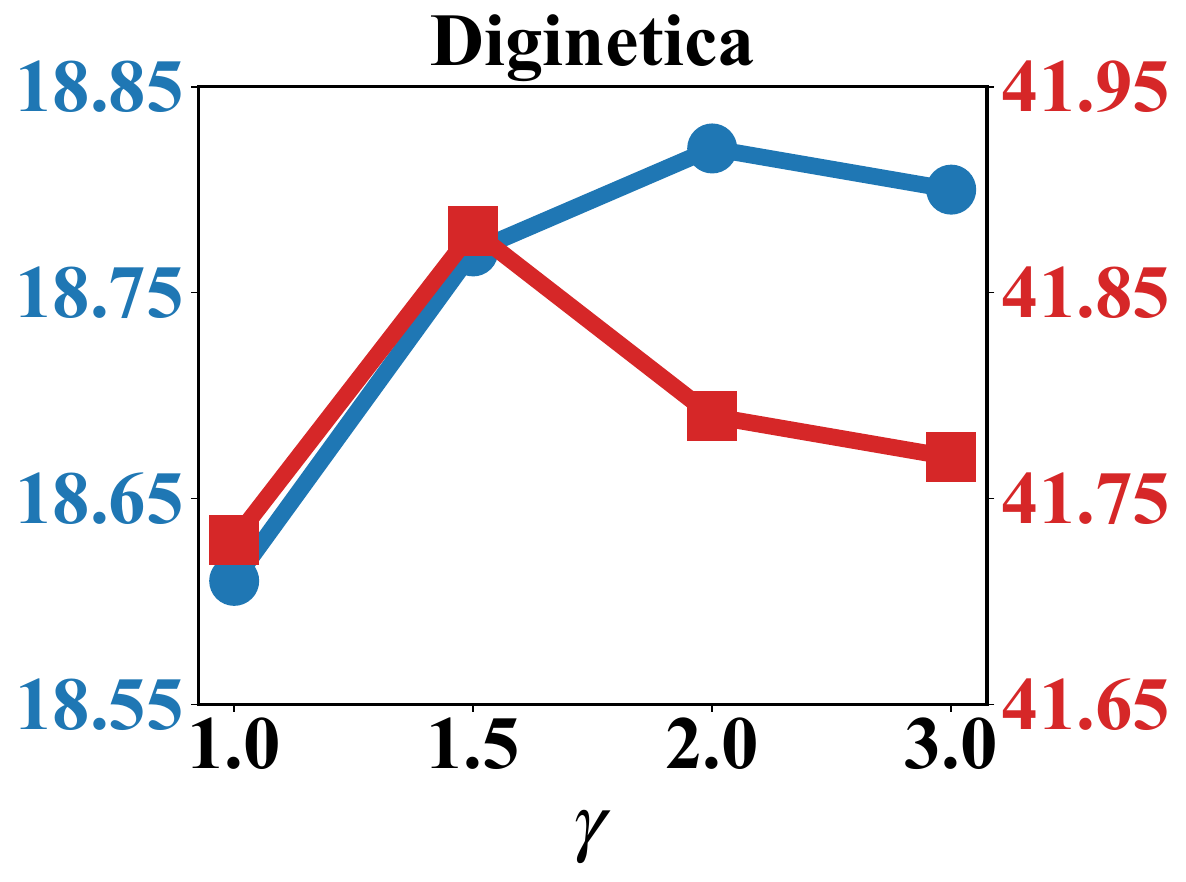}
\includegraphics[width=0.25\linewidth]{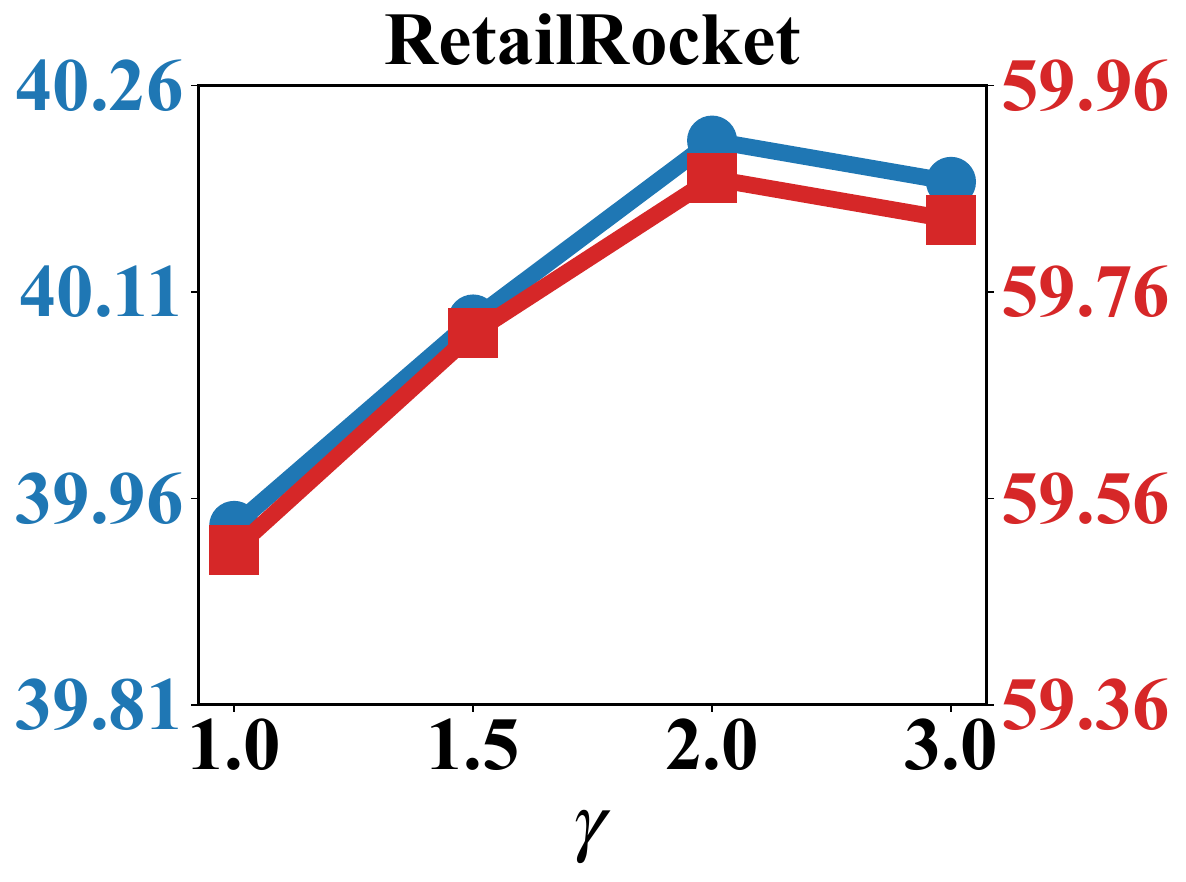}
\includegraphics[width=0.25\linewidth]{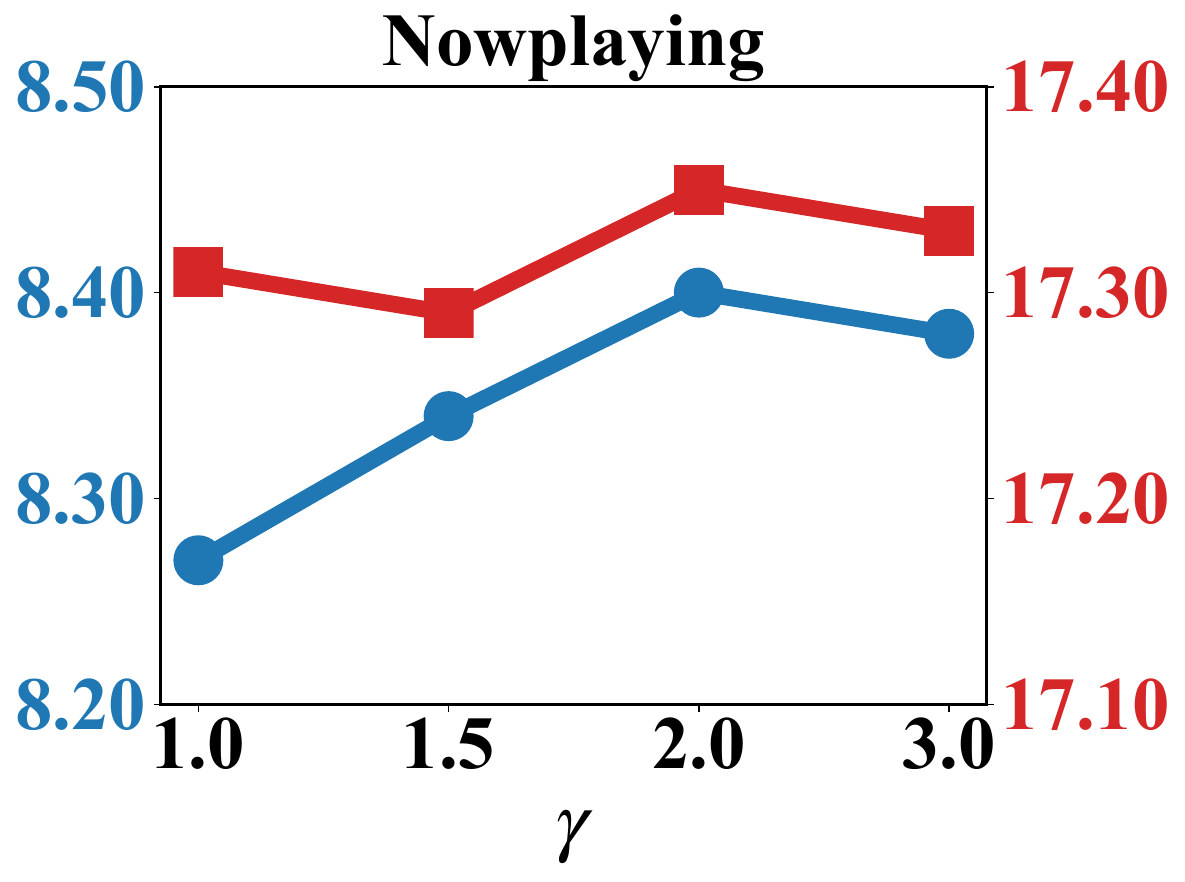}
\caption{Hyperparameter sensitivity of TRUST with respect to the temporal reliability decay exponent $\gamma$. HR@10 and MRR@10 are reported on Diginetica, RetailRocket, and Nowplaying.}
\label{fig:hyper-gamma}
\end{figure*}

The decay exponent $\gamma$ in Eq.~\eqref{eq:uncertainty-temporal-reliability} controls how sharply the reliability score decreases when an observed interval deviates from the item-specific intervals reference. Figure~\ref{fig:hyper-gamma} shows that TRUST performs consistently across the tested range $\gamma\in\{1.0,1.5,2.0,3.0\}$. On all three datasets, MRR@10 improves from $\gamma=1.0$ to $\gamma=2.0$, indicating that a Gaussian-like decay provides a useful balance between robustness and discrimination. When $\gamma$ is further increased to $3.0$, performance becomes slightly lower or nearly unchanged, suggesting that an overly sharp decay may over-penalize moderately atypical but still informative intervals. HR@10 follows a similar but less monotonic pattern, with the best values occurring around $\gamma=2.0$. Therefore, we choose $\gamma=2.0$ for three datasets.

\subsection{RQ5: Algorithm Efficiency}
\label{sec:rq5}

Last but not least, we ask whether our proposed method improves performance at the cost of high computational overhead.
We compare TRUST with representative baselines on actual per-epoch training time and the number of convergence epochs (Table~\ref{tab:efficiency}). TRUST adds two costs relative to SBRs: (i) one-time offline pre-computation of ITSF, and (ii) Multi-granularity of PIA. The precomputation step takes 37 seconds and runs once before training. During training, the ITSF reduces to a vectorized binary search over pre-sorted arrays, so ITSF does not require iterative parameter learning.

Table~\ref{tab:efficiency} further shows that TRUST takes longer per epoch than most temporal baselines and converges after more epochs on Diginetica. The full architecture has a moderate overhead in exchange for improvement of performance. ITSF pre-computation is small compared with model training, but PIA makes the model heavier than the last item-based session generation TSBRs.

\begin{table}[thb]
\centering
\caption{Actual running time on Diginetica.}
\label{tab:efficiency}
\setlength{\tabcolsep}{5pt}
\begin{tabular}{lcc}
\toprule
Method & Time/epoch (s) & Conv.\ epochs \\
\midrule

DMI-GNN   & 221 & 12            \\
IGT       & 323 & \phantom{0}9  \\
TE-GNN    & 182 & 10            \\
TMI-GNN   & 209 & 10            \\
DT-GAT    & 488 & \phantom{0}8  \\
\midrule
TRUST     & 439 & 14 \\
\phantom{0}+ ITSF precomp. &   37 & --- \\
\bottomrule
\end{tabular}
\end{table}




\section{Conclusion}

This paper presents TRUST, a temporal reliability calibration framework for session-based recommendation. TRUST calibrates the temporal interval as item-specific evidence whose meaning depends on the empirical interval pattern of the corresponding item. Experiments across three datasets show that this calibration improves recommendation performance and can also benefit several existing temporal SBR backbones. Future work can extend this line of work with tail-aware and shift-aware calibration strategies. The sample-size-dependent margin in ITSF improves the robustness of item-level temporal calibration for sparsely observed items. Future work can build on this design by incorporating one-sided tail calibration and category-level temporal references to further enhance temporal reliability estimation in long-tail recommendation scenarios.

\bibliographystyle{IEEEtran}
\bibliography{references}

\appendices
\section{Baseline Descriptions}
\label{app:baselines}

\noindent\textbf{GRU4Rec}~\cite{hidasi2015session} introduces recurrent neural networks into session-based recommendation using GRU units to model intra-session sequential patterns for next-item prediction.

\noindent\textbf{NARM}~\cite{li2017neural} extends recurrent session modeling with an attention mechanism to jointly represent the user's main intent and more localized sequential preference within a session.

\noindent\textbf{SR-GNN}~\cite{wu2019session} formulates each session as a directed graph and applies gated graph neural networks to learn item representations, followed by an attention-based readout.

\noindent\textbf{FGNN}~\cite{qiu2019rethinking} enhances graph-based session modeling with richer interaction structures to capture item dependencies from both sequential transitions and graph relations.

\noindent\textbf{GCE-GNN}~\cite{wang2020global} augments session-level graph modeling with global item co-occurrence information, allowing local session representations to benefit from broader collaborative context.

\noindent\textbf{$S^2$-DHCN}~\cite{xia2021selfhypergraph} models high-order item and session relationships through hypergraph convolution and introduces self-supervised learning to improve representation quality.

\noindent\textbf{MSGAT}~\cite{qiao2023bi} applies sparse graph attention to reduce noise in item-level and session-level representations for more robust graph-based session modeling.

\noindent\textbf{DMI-GNN}~\cite{lv2025dynamic}: introduces multi-interest learning into session modeling, uses multiple positional patterns to encode different positional contexts, and applies dynamic multi-interest regularization to reduce redundant interest representations based on session length.

\noindent\textbf{STAN}~\cite{garg2019sequence} adapts neighborhood-based session recommendation with temporal decay, assigning lower importance to distant sessions and higher importance to recent interactions.

\noindent\textbf{TASRec}~\cite{zhou2021temporal} incorporates historical sessions across different days using a temporally decaying mechanism to assign adaptive importance to past user-interest evidence.

\noindent\textbf{TE-GNN}~\cite{tang2022time} incorporates time-enhanced item transition information into graph neural networks, improving session representation by modeling temporally dependent interaction patterns.

\noindent\textbf{IGT}~\cite{wang2023interval} explicitly exploits inter-item intervals within a session to refine item relation modeling and improve session representation construction.

\noindent\textbf{TMI-GNN}~\cite{shen2023temporal} introduces temporal-aware multi-interest modeling to capture evolving user preferences from time-sensitive item transitions within a session.

\noindent\textbf{DT-GAT}~\cite{guo2026dual} uses time-aware graph and hyper-graph channels to incorporate interaction intervals and inter-session temporal differences into item-level and session-level representation learning.

\end{document}